\documentclass[a4paper,11pt]{article}
\usepackage{jheppub} 
\usepackage{lineno}
\usepackage{physics}
\usepackage{cancel}
\usepackage{feynmp}
\usepackage{appendix}
\usepackage{graphicx}
\usepackage{caption}
\usepackage{subcaption}
\usepackage{svg}
\usepackage{multirow}
\usepackage{multicol}
\usepackage{makecell}
\usepackage{footnote}
\usepackage{mathtools}
\usepackage{orcidlink}

\def\D{\mathrm{D}}

\title{\boldmath Feynman Integral Reduction using Syzygy-Constrained Symbolic Reduction Rules}

\author[\!a,b,c]{Sid Smith,}\emailAdd{sid.smith@ed.ac.uk}

\author[\,a]{Mao Zeng}\emailAdd{mao.zeng@ed.ac.uk}

\affiliation{$^a$Higgs Centre for Theoretical Physics, University of Edinburgh, James Clerk Maxwell Building,Peter Guthrie Tait Road, Edinburgh, EH9 3FD, United Kingdom}
\affiliation{$^b$Dipartimento di Fisica e Astronomia, Universita di Padova, Via Marzolo 8, 35131 Padova, Italy}
\affiliation{$^c$INFN, Sezione di Padova,
Via Marzolo 8, I-35131 Padova, Italy.}

\abstract{We present a new algorithm for integration-by-parts (IBP) reduction of Feynman integrals with high powers of numerators or propagators, a demanding computational step in evaluating multi-loop scattering amplitudes.
The algorithm starts with solving syzygy equations in individual sectors to produce IBP operators that turn seed integrals into IBP equations without artificially raised propagator powers. The IBP operators are expressed in terms of index-shift operators and number operators. We perform row reduction to systematically reshuffle the IBP operators and expose reduction rules with symbolic dependence on the powers of propagators and numerators. When this is insufficient, we produce more symbolic reduction rules by directly solving the linear system of IBP equations in which some propagator/numerator powers are kept symbolic. This linear system is kept small, as the equations are generated from a small set of seed integrals in the neighborhood of the target integral. We stress-test our algorithm against two highly non-trivial examples, namely rank-20 integrals for the double box with an external mass and the massless pentabox. As an application, we revisit the IBP reduction in a calculation of scattering amplitudes for spinning black hole binary systems, which involves two-loop Feynman integrals with complexity greater than 20, and achieve much faster IBP reduction than that of the original calculation.}

\begin{document}
\maketitle
\flushbottom
\begin{fmffile}{feyndiagrams}

\newpage

\section{Introduction}

Scattering amplitudes are the fundamental objects within perturbative quantum field theories (QFTs), and allow us to bridge the gap between theoretical predictions and experimental data. Precision calculations of scattering amplitudes have applications in collider physics, gravitational wave physics, cosmology and many other fields. In many calculations, there are a large number of complex, multi-loop Feynman integrals to evaluate, and integration-by-parts (IBP) reduction \cite{Chetyrkin:1981qh} is used to rewrite these Feynman integrals as linear combinations of a minimal set of integrals of a lower complexity, known as master integrals (MIs) \cite{Smirnov:2010hn,Lee:2013hzt}. 

IBP reduction is commonly carried out using the Laporta algorithm \cite{Laporta:2000dsw}, which has been implemented in many publicly available programs, such as AIR \cite{Anastasiou:2004vj}, LiteRed \cite{Lee:2012cn,Lee:2013mka}, FIRE \cite{Smirnov:2008iw,Smirnov:2013dia,Smirnov:2014hma,Smirnov:2019qkx,Smirnov:2023yhb}, Reduze \cite{Studerus:2009ye,vonManteuffel:2012np} and Kira \cite{Maierhofer:2017gsa,Maierhofer:2018gpa,Klappert:2020nbg,Lange:2025fba}. Despite the efficiency of these programs, the IBP reduction step frequently remains the most computationally heavy step of the calculation, due to the size of the system of equations becoming larger and larger for more complex integrals. Because of this, there have been many developments in recent years in the hope to reduce the size of this system, one involving the introduction of methods from algebraic geometry \cite{Tarasov:2004ks,Smirnov:2005ky,Smirnov:2006tz,Smirnov:2006wh,Gerdt_2006,Lee:2008tj,Barakat:2022qlc,Bendle:2019csk,Ita:2015tya,Larsen:2015ped,Gluza:2010ws}, notably the inclusion of syzygy equation constraints to reduce the number of unphysical auxiliary integrals in the system. These methods have been implemented in the computer program NeatIBP \cite{Wu:2023upw,Wu:2025aeg}. Another idea for speeding up IBP reduction is re-organizing IBP equations in block triangular forms, as implemented in Blade \cite{Guan:2024byi}. There have also been explorations made into other methods to perform this reduction that bypass the use of IBPs, such as intersection theory \cite{Mastrolia:2018uzb,Frellesvig:2019kgj,Frellesvig:2019uqt,Frellesvig:2020qot,Chestnov:2022xsy,Fontana:2023amt,Brunello:2023rpq}.

With the recent development of finite field numerical techniques \cite{Kant:2013vta,vonManteuffel:2014ixa,Peraro:2016wsq,Abreu:2018zmy,Klappert:2019emp,Peraro:2019svx,Laurentis:2019bjh,DeLaurentis:2022otd,Magerya:2022hvj,Belitsky:2023qho,Chawdhry:2023yyx,Liu:2023cgs,Maier:2024djk} to reconstruct analytic expressions (in the spacetime dimension and kinematic invariants) using a large number of numerical evaluations, it is important to make each numerical IBP reduction run as fast as possible. However, a significant computational challenge in both Laporta and syzygy approaches is solving large linear systems of IBP equations. A closely related challenge is finding good heuristic choices of ``seed integrals'', essentially anchor points controlling which IBP equations are generated. These choices are generally based on experience rather than any underlying principles. Recent literature has explored improved heuristic seeding choices \cite{Driesse:2024xad, Guan:2024byi, Bern:2024adl, Lange:2025fba} as well as AI techniques for this problem \cite{vonHippel:2025okr,Song:2025pwy,Zeng:2025xbh}. These challenges are bypassed when one produces symbolic reduction rules that directly reduce arbitrary target integrals to lower-complexity integrals; such rules can be applied iteratively until all integrals are reduced to master integrals. This is sometimes referred to as ``parametric reduction'', as in a four-loop application in Ref.~\cite{Ruijl:2017cxj}. A public package for this approach is LiteRed \cite{Lee:2013mka}. Refs.~\cite{Guan:2023avw, Hu:2024kch} explored generating function methods for finding reduction rules for high-complexity integrals. Direct solutions of IBP equations generated from syzygy equations have been studied in Ref.~\cite{Kosower:2018obg}. Despite these advances, the use of symbolic reduction rules has found limited applications in QFT calculations relative to e.g.\ the wide use of the Laporta algorithm. Specifically, the use of symbolic reduction rules for multi-scale Feynman integrals beyond one loop is largely unexplored territory.

In this paper, we present a novel algorithm to generate symbolic reduction rules that can be applied to a specific set of target integrals. The algorithm combines the methods of spanning cuts \cite{Larsen:2015ped}, syzygy equation constraints, smart seeding, and judicious reshuffling of the identities, in order to generate these rules.
A main feature of our approach is that we aim to keep the explicit dependence of the reduction rules on the propagator/numerator powers $n_{i}$ minimal, by performing Gaussian elimination at the level of IBP operators in a way that is agnostic to the seed integrals used.\footnote{Similar operator-level reshuffling, though in the Laporta approach without syzygy equations, was silently added in the version 6.5 of FIRE \cite{Smirnov:2023yhb} and improved in a yet unreleased version \cite{Bern:2024adl}. The aim there is simplifying the IBP operators before normal seeding and linear solving, rather than directly finding reduction rules.} For a subset of Feynman integrals that cannot be reduced by the above rules, we additionally solve a linear system of equations formed by a small number of seed integrals in the neighborhood of the targeted integrals, to produce extra symbolic reduction rules to complete the IBP reduction. \footnote{In practice, this subset of integrals may depend on a number of factors, such as the monomial ordering chosen in the syzygy computation, or the ordering of the indices used to define the weight of an integral. There could potentially exist an optimal choice for each specific case, which is a problem that would require further exploration.}
We are able to use these rules to reduce integrals with multiple powers of different ISPs for highly nontrivial multi-scale integral topologies such as the two-loop pentabox, as we show in the examples.

\section{Integration-by-Parts: An Overview \label{sec:IBP}}

In this section we will give an overview of the method of integration-by-parts (IBPs) and summarise some recent developments which have had a significant effect on the state-of-the-art for this method. 

\subsection{Integration-by-Parts Identities}

We start by considering an arbitrary $L$-loop Feynman integral of the form
\begin{equation}
    F(n_{1},\dots,n_{N}) = \int\left(\prod_{b=1}^{L}\dd^{\D}\ell_{b}\right)\frac{1}{\rho_{1}^{n_{1}}\dots\rho_{N}^{n_{N}}}\,, \label{eq:family}
\end{equation}
where $\rho_{i}$ are known as the propagators, and are given in terms of scalar products involving the loop momenta $\ell_{b}$ and a set of external momenta $p_{i}$, $i=1,\dots,E$. The \textit{rank} of an integral refers to the total power of all numerators, i.e. negative powers under the fraction sign,
\begin{equation}
  \sum_i \theta(-n_i) (-n_i) \, ,
\end{equation}
and the number of \textit{dots} refers to the total number of excess denominator powers beyond $1$,
\begin{equation}
  \sum_i \theta(n_i) (n_i-1) \, .
\end{equation}
In this paper, complex integrals refer to integrals with a large number of dots or a high rank.

Eq.~\eqref{eq:family} with all possible index values $n_i$ defines a \textit{family} of Feynman integrals, and we aim to write any arbitrary Feynman integral in a given family as a linear combination of master integrals $J_{i}$
\begin{equation}
    F(\vec n) = \sum_{i}c_{i}J_{i}\,.
\end{equation}
We can use IBPs to find the coefficients $c_{i}$ by starting with the identity
\begin{equation}
    0 = \int\left(\prod_{b=1}^{L}\dd^{\D}\ell_{b}\right)\pdv{}{\ell_{a}^{\mu}}\left(\frac{q_{\alpha}^{\mu}}{\rho_{1}^{n_{1}}\dots\rho_{N}^{n_{N}}}\right)\,, \label{eq:ibp}
\end{equation}
where $q_{\alpha}^{\mu}\in\left\{\ell_{1},\dots,\ell_{L},p_{1},\dots,p_{E}\right\}$. By choosing all possible values of $a$ and $\alpha$, we have $L(L+E)$ identities.

In the Laporta algorithm \cite{Laporta:2000dsw}, different values of $\vec n$, known as the \textit{seed integrals}, are chosen under certain criteria and used in Eq.~\eqref{eq:ibp} to generate a system of equations. By solving this linear system of equations, any integrals of the family can be reduced to linear combinations of the master integrals. However, as we have mentioned earlier, the Laporta algorithm is by no means the only systematic way to perform IBP reduction.

\subsection{Sectors and Spanning Cuts}

In this paper, the sector of an integral is fully described by the positive powers of the indices $n_{i}$ in that integral. Therefore, for a given integral $F(\vec n)$, the corresponding sector is
\begin{equation}
    \vec m = \vec n\eval_{-\text{ve}\rightarrow0}\,.
\end{equation}
For example, the integral $F(2,1,1,-1,-2)$ belongs to the sector $(2,1,1,0,0)$. This is different to the standard definition of a sector, which is only concerned with which propagators have a positive power, not the specific value they take.

Given two sectors $\vec m_{1}$ and $\vec m_{2}$, we define $\vec m_{2}$ to be a \textit{subsector} of $\vec m_{1}$ if and only if
\begin{align}
    m_{1,i}\geq m_{2,i} \quad \forall \ i\,.
\end{align}
As we will see later, when we generate IBP identities within a given sector with syzygy constraints, only integrals within that sector or subsectors of it will appear in these relations.

In a given Feynman integral, some sectors will vanish in dimensional regularization, and this is usually due to the integrals in that sector being scaleless integrals. If we are to find all non-vanishing sectors, we are able to figure out which denominators must be present in order to obtain a full reduction, which will allow us to specify a set of \textit{spanning cuts}.

A cut, loosely speaking, is equivalent to setting a propagator to be on shell
\begin{equation}
    \frac{1}{\rho}\rightarrow\delta(\rho)\,.
\end{equation}
This is less trivial for squared propagators, but can be achieved through IBPs to write the squared propagator in terms of the unsquared ones. On a cut $C$, a Feynman integral will potentially vanish under the condition
\begin{equation}
    F_{C}(n_{1},\dots,n_{N}) = \begin{cases}
        F_{C}(n_{1},\dots,n_{N}), \quad n_{i}>0, \ \forall \ i\in C\,.\\
        0 \quad\qquad\qquad\qquad \ \ \text{otherwise}\,.
    \end{cases}
\end{equation}
The benefit of using cuts is that the coefficients in IBP equations are, in general \footnote{There are cases, where IBP equations do not commute with the cut operation, and one finds inconsistent equations across the cuts due to so-called \textit{magical relations} \cite{}. For the examples we consider in this paper, these relations do not appear, and in cases where they do, it is advised to use this algorithm without cuts.}, unaffected by them
\begin{equation}
    F = \sum_{i}c_{i}J_{i}\Rightarrow F_{C} = \sum_{i}c_{i}J_{C,i}\,,
\end{equation}
so if we can work out the IBP coefficients $c_{i}$ on a cut, we have them generically. By considering a system of IBP equations on a cut, we remove a significant number of variables and equations from our system, making it easier to solve.

Obviously some master integrals will vanish on a cut $C$, so it is
important that one chooses a spanning set of cuts that allows all
master integrals to appear at least once, in order to get every IBP
coefficient that exists. In practice, one can choose this set by
making sure that one includes all non-vanishing sectors in their list
of cuts. By solving the IBP equations on each of the spanning cuts and
combining the reductions together, we can obtain the full
reduction. The method of spanning cuts \cite{Larsen:2015ped} has been
automated in NeatIBP \cite{Wu:2023upw,Wu:2025aeg}, however almost all
codes
\cite{Anastasiou:2004vj,Lee:2012cn,Lee:2013mka,Smirnov:2008iw,Smirnov:2013dia,Smirnov:2014hma,Smirnov:2019qkx,Smirnov:2023yhb,Studerus:2009ye,vonManteuffel:2012np,Guan:2024byi}
allow various ways to cut propagators, allowing one to use this method
manually. Although for some of the examples listed in Section
\ref{sec:ex}, we use the spanning cuts approach, this is not a
requirement for the algorithm, which works independent of the use of
cuts.

\subsection{Building Identities from Syzygy Constraints}

We can reduce the size of the system generated by using syzygy constraints \cite{Tarasov:2004ks,Smirnov:2005ky,Smirnov:2006tz,Smirnov:2006wh,Gerdt_2006,Lee:2008tj,Barakat:2022qlc,Bendle:2019csk,Ita:2015tya,Larsen:2015ped,Gluza:2010ws} when building the identities. We can start with a more generic IBP identity of the form
\begin{equation}
    0 = \int\left(\prod_{b=1}^{L}\dd^{\D}\ell_{b}\right)\pdv{}{\ell_{a}^{\mu}}\left(\frac{P_{a\alpha}(\rho)q_{\alpha}^{\mu}}{\rho_{1}^{n_{1}}\dots\rho_{N}^{n_{N}}}\right)\,,
\end{equation}
where $P_{a\alpha}(\rho)$ are polynomials in the propagators. This can actually be seen as a linear combination of Laporta-style identities Eq.~\eqref{eq:ibp} with different values of the seed integrals $\vec n$, and therefore no information is lost by this. We are now going to enforce the following constraints
\begin{equation}
    P_{a\alpha}(\rho)q_{\alpha}^{\mu}\pdv{}{\ell_{a}^{\mu}}\rho_{i} = f_{i}(\rho)\rho_{i}, \quad \forall\, i\in\sigma\,,
\end{equation}
on a subset $\sigma$ of the propagators. Typically $\sigma$ is chosen to be all the propagators that appear as a denominator in the given sector
\begin{align}
    \sigma = \{i\,|\,n_{i}>0\}\,.
\end{align}
This constraint means that the identities we generate will never contain integrals with powers larger than $n_{i}$ for the denominators. So for example if you input a seed integral with $n_{i}=1$, and $i\in\sigma$, the resulting equations will never contain an integral with $n_{i}>1$. The benefit of this is that we can avoid including variables in our system that are unnecessary for a specific target integral.

In order to find these polynomials $P_{a\alpha}(\rho)$, we first specify the set of $R=L(L+E)$ IBP generators and polynomials  
\begin{equation}
    G_{i}\in\left\{q_{1}^{\mu}\pdv{}{\ell_{1}^{\mu}},\dots,q_{L+E}^{\mu}\pdv{}{\ell_{L}^{\mu}}\right\}, \quad P_{i}(\rho)\in\left\{P_{11}(\rho),\dots,P_{L,L+E}(\rho)\right\}\,,
\end{equation}
and the number of propagators that are constrained $T=\abs{\sigma}$. We can then formulate our condition as
\begin{equation}
    \vec c^{T}M = 0\,,
\end{equation}
where the vector $\vec c$ has $R+T$ elements
\begin{equation}
    c_{i} = \begin{cases}
        P_{i},  &i\leq R\,,\\
        f_{i-R}(\rho),  &\text{otherwise}\,,
    \end{cases}
\end{equation}
The $(R+T)\times T$ matrix $M$ is defined as
\begin{equation}\label{eq:module}
    M_{ij} = \begin{cases}
        G_{i}(\rho_{\sigma(j)}), &i\leq R\\
        -\rho_{\sigma(j)}\delta_{i-R,j}, &\text{otherwise}
    \end{cases}
\end{equation}

The matrix $M$ is also interpreted as a module over the polynomial ring $\mathbb{Q}(\{x\})[\rho_1,\rho_2,\dots,\rho_N]$ with rational coefficients of the kinematic invariants $\{x\}$, generated by the row vectors of the matrix. Then the above equation becomes a \emph{syzygy equation}, in the language of computational algebraic geometry, over this module $M$. One can use programs such as Singular \cite{DGPS} to find the solutions $\vec c$ to this syzygy equation. These solutions give us choices we can make for the polynomials $P_{a\alpha}(\rho)$, allowing us to build identities without unnecessarily raising propagator powers.

We can also generate syzygy-constrained identities on a cut $C$ as well, assuming that all these cut particles appear as single-power propagators, our Feynman integral takes the form
\begin{equation}
    F_{C}(\vec{n}) = \int\left(\prod_{b=1}^{L}\dd^{\D}\ell_{b}\right)\left(\prod_{i\notin C}\rho_{i}^{-n_{i}}\right)\left(\prod_{i\in C}\delta(\rho_{i})\right)\,.
\end{equation}
In this case, our generic IBP identity will look like
\begin{equation}
    0 = \int\left(\prod_{b=1}^{L}\dd^{\D}\ell_{b}\right)\pdv{}{\ell_{a}^{\mu}}\left(P_{a\alpha}(\rho)q_{\alpha}^{\mu}\prod_{i\notin C}\rho_{i}^{-n_{i}}\prod_{i\in C}\delta(\rho_{i})\right)\,.
\end{equation}
Therefore we just have to ensure that the cut propagators are also constrained in the syzygy equation to make sure the action of the operator on the $\delta$-functions vanishes. But these syzygy equations are simpler as we are free to set the cut propagators to zero, therefore we just have the change
\begin{equation}
    M_{C} = M\eval_{\rho_{i}\rightarrow0,\,i\in C}\,.
\end{equation}

\section{Algorithm for Reducing Complex Integrals \label{sec:algorithm}}

\begin{figure}[h!]
    \centering
    \begin{fmfgraph*}(200,100)
        \fmfforce{(0,0.5h)}{i1}
        \fmfforce{(w,0.5h)}{i2}
        \fmfforce{(0.3w,0.5h)}{v1}
        \fmfforce{(0.7w,0.5h)}{v2}
        \fmf{fermion,label=$p$}{i1,v1}
        \fmf{plain}{v2,i2}
        \fmf{fermion,left,label=$\ell_{1}$}{v1,v2}
        \fmf{plain,left}{v2,v1}
        \fmf{fermion,label=$\ell_{2}$,label.side=left}{v1,v2}
    \end{fmfgraph*}
    \caption{Two Loop Equal Mass Sunrise \label{fig:Sunrise}}
\end{figure}
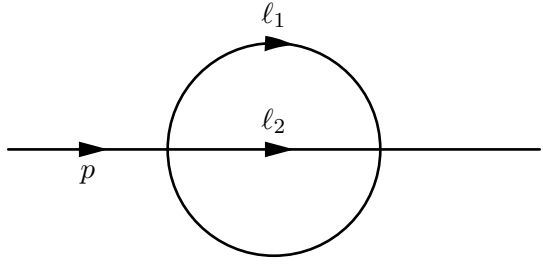

In this section we describe our algorithm for reducing a set of integrals of arbitrarily high complexity. The algorithm is split into two main stages
\begin{enumerate}
    \item Building a complete set of reduction rules that will be able to reduce any arbitrary integral.
    \item Applying these reduction rules to reduce a specific set of target integrals to master integrals.
\end{enumerate}
We will go through each of these stages in detail, using the equal mass sunrise to illustrate each step clearly, as shown in Fig. \ref{fig:Sunrise}.

This topology has a propagator set
\begin{equation}
    \begin{aligned}
    \rho_{1} = \ell_{1}^{2}-m^{2}, \quad &\rho_{2} = \ell_{2}^{2}-m^{2}, \quad \rho_{3} = (\ell_{1}+\ell_{2}-p)^{2}-m^{2}\\
    \rho_{4} &= \ell_{1}\cdot p, \quad \rho_{5} = \ell_{2}\cdot p\,,
    \end{aligned}
\end{equation}
where $\rho_{4}$ and $\rho_{5}$ are ISPs, and $p^{2}=s$. We will be reducing the following set of target integrals
\begin{equation}
    \{F(2,1,1,-4,-4),F(1,2,1,0,-7),F(1,1,1,-6,-4),F(1,1,1,-11,0)\}\,.\label{eq:target}
\end{equation}
Further details of the various steps of this algorithm are given in Appendix \ref{app:Example}.

\subsection{Building Reduction Rules}

\begin{figure}[h!]
    \centering
    \includegraphics[scale=0.7]{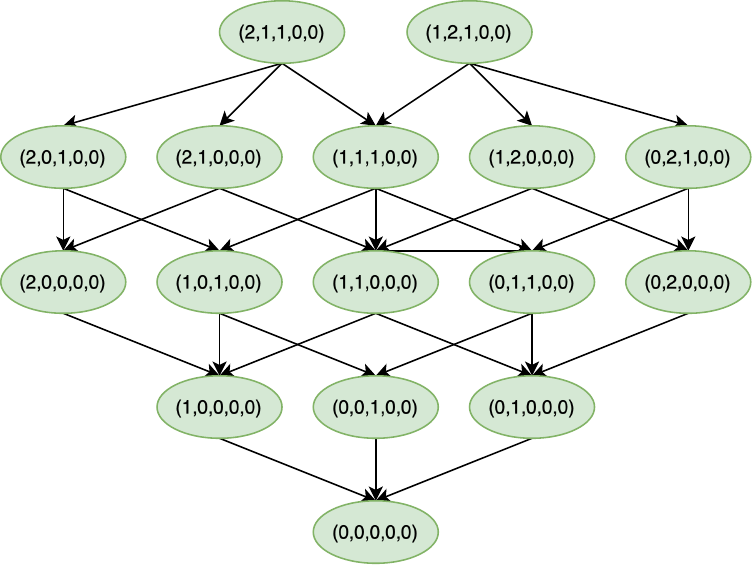}
    \caption{Tower of Sectors one must consider when the top sectors are $(2,1,1,0,0)$ and $(1,2,1,0,0)$. This drawing is schematic and the horizontal levels here do not correlate with the weighting described later. The relevant information is contained within the arrows, denoting the subsector inheritance hierarchy.\label{fig:sectors}}
\end{figure}

Our algorithm builds identities on each individual sector one at a time, using the definition of sector described in Section \ref{sec:IBP}. The sectors associated to each of our target integrals in Eq. \ref{eq:target} respectively are
\begin{equation}
    \{(2,1,1,0,0),(1,2,1,0,0),(1,1,1,0,0),(1,1,1,0,0)\}\,.
\end{equation}
From this we can remove duplicates and subsectors to obtain the set of \textit{top sectors}
\begin{equation}
    \{(2,1,1,0,0),(1,2,1,0,0)\}\,.
\end{equation}
We can then determine our tower of sub-sectors, as shown in Fig. \ref{fig:sectors}. It is clear that in this case, many of these sectors are zero sectors, and these can be ignored as all integrals in these sectors are zero.

\begin{figure}[h!]
    \centering
    \includegraphics[scale=0.7]{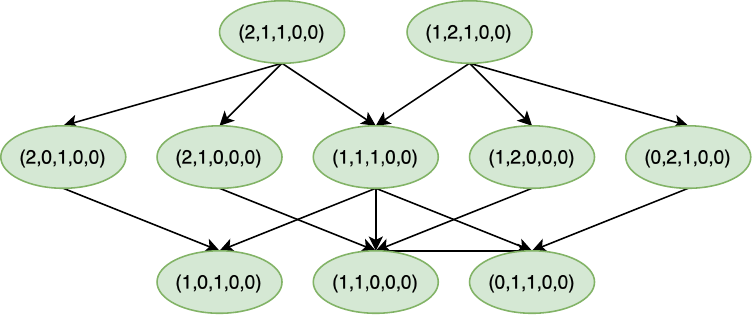}
    \caption{Tower of Non Zero Sectors for this Example.\label{fig:nonzerosectors}}
\end{figure}

Within each non-zero sector $\vec m$ on this tower, we will build reduction rules to reduce any arbitrary integral on this sector. In this case we only have $10$ sectors to consider, as shown in Fig. \ref{fig:nonzerosectors}.

In this paper, we will only describe the details of the algorithm for the sectors that are independent up to symmetries. For example, there is the symmetry
\begin{equation}
    (2,1,1,0,0)\xleftrightarrow{\ell_{1}\leftrightarrow\ell_{2}}(1,2,1,0,0)\,.
\end{equation}

In fact, up to such symmetry relations, there are only $4$ independent sectors
\begin{equation}
    \{(2,1,1,0,0),(1,1,1,0,0),(2,1,0,0,0),(1,1,0,0,0)\}\,.
\end{equation}
This organisation of sectors is similarly done in NeatIBP \cite{Wu:2023upw,Wu:2025aeg}.

In order to reduce any integral, there needs to be a notion of what it means to reduce the integral, and for this we have to specify how to determine the \textit{weight} of an integral\footnote{The term weight here is not to be confused with the identical term used to describe polylogs, these are completely different concepts}. We use the following weight function
\begin{equation}
    W(\vec n) = \left(\sum_{n_{i}>0},\sum_{n_{i}>0}(n_{i}-1),-\sum_{n_{i}<0}n_{i},-\mathcal{O}(\abs{\vec n})\right)\,.
\end{equation}
where $\mathcal{O}$ is some arbitrary ordering of the indices. 

To compare the weight of two integrals, compare the first element of this weight function, if they are equal then compare the second element, and so on until there is a difference. If there is no difference then these integrals are identical.

In this example, the most optimal way to choose the weight function is
\begin{equation}
    \mathcal{O}(\abs{\vec n}) = (\abs{n_{1}},\abs{n_{2}},\abs{n_{3}},\abs{n_{5}},\abs{n_{4}})\,,
\end{equation}
as this order makes our target integrals lower weight.

\paragraph{Generating Ordered Identities}

In each sector $\vec m$, we start off by solving the syzygy equations with constraints on the subset $\sigma=\{i\,|\,m_{i}>0\}$ of propagators
\begin{equation}
    P_{a\alpha}(\rho)q_{\alpha}^{\mu}\pdv{}{\ell_{a}^{\mu}}\rho_{i} = f_{i}(\rho)\rho_{i}, \quad \forall\, i\in\sigma\,.
\end{equation}
The module matrix is constructed from Eq. \ref{eq:module}, and is given in Appendix \ref{app:Example}.

From our solutions, our identities will take the form
\begin{equation}
    0 = \sum_{i}(\alpha_{i}+\vec{\beta}_{i}\cdot\vec{n})F(\vec{n}+\vec{\gamma}_{i})\,.
\end{equation}
Since we are fully working in a sector $\vec m$, we fix some indices for the sector that we are in a priori
\begin{equation}
    n_{i} = \begin{cases}
        m_{i},& i\in\sigma\,,\\
        \eta_{i},& i\notin\sigma\,.
    \end{cases}\\
\end{equation}
where $m_{i}$ are fixed numbers, and $\eta_{i}$ are still free to be
anything. This is equivalent to shifting
$\alpha_{i}\rightarrow\alpha_{i}+\vec\beta_{i}\cdot\vec m$ and
shortening $\vec\beta_{i}$ to only contain the non-sector
indices. This constitutes the main motivation for using syzygy
constraints in our reduction rules. If we were using Laporta
identities we would have to consider all the indices to be free. Said
another way, using syzygy constraints removes many degrees of freedom
to be resolved from our system.

We also can remove any identities that do not contain any integrals in the sector we are working on. The criterion for this is if an identity does not contain any $\vec\gamma_{i}$ such that
\begin{equation}
    (\vec\gamma_{i})_{j}=0 \quad \forall\,j\in\sigma\,,
\end{equation}
then it is discarded.\footnote{A similar technique is also used in NeatIBP \cite{Wu:2025aeg}, referred to as the maximal cut method. The difference is that they do this at the syzygy solution level, by finding which syzygies contribute to the Gr\"obner basis of the maximal cut syzygies.} Note that the identities do not raise propagator powers, so any shift of a propagator power is negative and leads to identities in subsectors. After this we are left with identities of the form
\begin{equation}
    0 = \sum_{i}(\alpha_{i}+\vec{\beta}_{i}\cdot\vec{\eta})F[\vec{n}+\vec{\gamma}_{i}]\,, \label{eq:ibpIdentities}
\end{equation}
For the sunrise example, some details are provided in Table~\ref{tab:tab1}.

\begin{table}
\centering
    \begin{tabular}{|c|c|c|}
        \hline
        Sector&Number of Syzygy Solutions&Number of Relevant Identities\\
        \hline
        (2,1,1,0,0)&36&13\\
        \hline
        (1,1,1,0,0)&36&6\\
        \hline
        (2,1,0,0,0)&12&9\\
        \hline
        (1,1,0,0,0)&12&6\\
        \hline
    \end{tabular} 
    \caption{Table showing the number of syzygy solutions found on each sector and the number of identities that are initially used. \label{tab:tab1}}
\end{table}

The final step is now to order these identities such that the first term is always the highest weight integral, so that they take the form of reduction rules. For this we have to introduce a sector-specific shift weight function
\begin{equation}
    w(\vec{\gamma}) = (\vec{\gamma}\cdot\vec{\xi},-\vec{\gamma}\cdot\vec{\theta},-\mathcal{O}(\vec{\gamma}))\,,
\end{equation}
where
\begin{equation}
    \vec{\xi} = \vec{m}\eval_{+\text{ve}\rightarrow1}, \quad \theta_{i} = 1-\xi_{i}\,,
\end{equation}
This weights shifts $\vec \gamma_i$ in Eq.~\eqref{eq:ibpIdentities}  according to first whether they take integrals out of their sector into subsectors, if they do they are considered to have the lowest weight. If it doesn't then it orders them according to the rank of the shift for the non-sector indices. If this is the same it will order them according to some specific ordering of the indices. The list of ordered identities for each of our sectors is given in Appendix \ref{app:Example}.

Because we have ordered these identities according to the weight of their shift, they are naturally in the form of reduction rules. We then analyse the highest weight shift vectors for each of these identities to see which ones are necessary, which are redundant, and what types of integrals cannot be reduced with these identities.

For example, on the sector $\vec m=(1,1,1,0,0)$, we find there are $6$ initial identities, the first of which takes the form
\begin{equation}
    2(2n_{4}-n_{5})F(1,1,1,n_{4},n_{5}-1)+\text{lower-weight integrals}=0\,.
\end{equation}
The full expression for this identity can be found in Eq. \ref{eq:fullId}. It is worth reiterating that $n_{4},n_{5}\leq0$ to ensure the resulting equation only contains integrals where these two indices act as ISPs. This identity can reduce all integrals in this sector except for those of the form
\begin{equation}
    F(1,1,1,n_{4},2n_{4}-1), \quad F(1,1,1,n_{4},0)\,.
\end{equation}
The first of these is not possible to reduce because of the coefficient in front of this leading weight term vanishes. The second integral cannot be reduced because it would involve inputting a seed of $n_{5}=1$, which would generate denominators for propagator $4$ in the other terms. Another identity has the leading weight term
\begin{equation}
    6s(2n_{4}-n_{5})F(1,1,1,n_{4},n_{5}-1)+\text{lower-weight integrals}=0\,,
\end{equation}
which is no longer helpful as its purpose is already being served by the previous identity we found. We find another identity with leading weight term
\begin{equation}
    4(1-\D+n_{5})F(1,1,1,n_{4},n_{5}-2)+\text{lower-weight integrals}=0\,,
\end{equation}
which is helpful as this allows us to reduce the integrals $F(1,1,1,n_{4},2n_{4}-1)$, because there is no situation where the coefficient of the leading weight term vanishes. However from both these identities we have included, we still cannot reduce integrals of the form
\begin{equation}
    F(1,1,1,0,-1), \quad F(1,1,1,0,0), \quad F(1,1,1,n_{4},0)\,.
\end{equation}
In fact these integrals cannot be reduced by any of the identities we have, and so we must move on to the next step for these. In Appendix $\ref{app:Example}$, we give a table showing exactly which identities are used and what they are unable to reduce.

\paragraph{Row Reducing Identities}

We move on to this step if we still have some arbitrary integrals that cannot be reduced from the previous step. In this step, we analyze the irreducible integrals, and fix whatever indices we are able to fix. For example, if we wanted to target the integral $F(1,1,1,n_{4},0)$, we could also fix $n_{5}=0$, and keep only one free index $n_{4}$.

Starting with the initial identities from before
\begin{equation}
    0 = \sum_{i}(\alpha_{i}+\vec{\beta}_{i}\cdot\vec{\eta})F(\vec{n}+\vec{\gamma}_{i})\,,
\end{equation}
with $\vec\eta$ representing the non-sector indices, we input seed integrals that correspond to small perturbations around the integrals we still haven't reduced. These perturbations have a maximum total negative value of $1$. For example for the target integral $F(1,1,1,n_{4},0)$, we input seed integrals
\begin{equation}
    \{(1,1,1,n_{4},0),(1,1,1,n_{4}-1,0),(1,1,1,n_{4},-1)\}\,.
\end{equation}
Putting this into the identities, we end up with $3$ times the number of identities as before, and we shorten the index vector $\vec\eta=(n_{4})$ to write them as
\begin{align}
    0 = \sum_{i}(\alpha_{ki}+\vec\beta_{ki}\cdot\vec\eta)F(\vec n+\vec\gamma_{ki})\,,
\end{align}
where $k=1,\dots,I$, and $I$ is the number of identities. We can then gather up all the $M$ shift vectors present and order them according to their weight, so that we can write these identities in matrix form
\begin{equation}
    \begin{pmatrix}
        (\alpha_{11},\vec\beta_{11})\cdot(1,\vec\eta)&(\alpha_{12},\vec\beta_{12})\cdot(1,\vec\eta)&\cdots&(\alpha_{1M},\vec\beta_{1M})\cdot(1,\vec\eta)\\
        (\alpha_{21},\vec\beta_{21})\cdot(1,\vec\eta)&(\alpha_{22},\vec\beta_{22})\cdot(1,\vec\eta)&\cdots&(\alpha_{2M},\vec\beta_{2M})\cdot(1,\vec\eta)\\
        \vdots&\vdots&\vdots&\vdots\\
        (\alpha_{I1},\vec\beta_{I1})\cdot(1,\vec\eta)&(\alpha_{I2},\vec\beta_{I2})\cdot(1,\vec\eta)&\cdots&(\alpha_{IM},\vec\beta_{IM})\cdot(1,\vec\eta)\\
    \end{pmatrix}\begin{pmatrix}
        F(\vec n+\vec\gamma_{1})\\
        F(\vec n+\vec\gamma_{2})\\
        \vdots\\
        F(\vec n+\vec\gamma_{M})
    \end{pmatrix} = 0\,.
\end{equation}
We are now free to isolate the matrix of coefficients
\begin{equation}
    \begin{pmatrix}
        \alpha_{11}&\vec\beta_{11}&\cdots&\alpha_{1M}&\vec\beta_{1M}\\
        \vdots&\vdots&\vdots&\vdots&\vdots\\
        \alpha_{I1}&\vec\beta_{I1}&\cdots&\alpha_{IM}&\vec\beta_{IM}
    \end{pmatrix}\,,
\end{equation}
and row reduce this matrix, then put it back into the matrix form of
the equations above to find equally valid identities. Similarly to the
previous section, we can then write these row reduced identities as
new reduction rules which can be used to reduce the previously
irreducible integrals. This row reduction is done using FiniteFlow,
and we only perform an analytic reconstruction on the rows
corresponding to reduction rules that we would actually need to
use. Since these identities depend only on the
  kinematic invariants, as well as abstract raising/lowering
  operators, this matrix will consist of rational functions of these
  invariants, not the propagator powers $n_{i}$.

We can perform this step iteratively, by fixing more indices after some free ones have been resolved, until we cannot find any more useful reduction rules this way. If we are still unable to resolve some integrals, we move on to the final step.

For example, for sector $\vec m=(1,1,1,0,0)$, this step does not allow us to reduce any of the leftover integrals, and so we have to take them all to the final step. In appendix \ref{app:Example} we describe this procedure in more detail.

Although this step did not give us any new reduction rules in this example, it is a crucial step for more complex topologies. In a lot of cases moving on to the final step is not needed, and even if it is, this step can reduce the number of unresolved integrals drastically, simplifying the final step.

\paragraph{Solving Directly for Missing Rules}

For the final step, we analyze the list of irreducible integrals, and attempt to find symbolic reduction rules for these in terms of the indices $n_{i}$. The way we do this is by generating a small system of equations around the fixed point specified by these integrals, i.e.\ using seed integrals in the neighborhoood of the targeted integrals, keeping the analytic dependence on the indices $n_{i}$. We then solve analytically using FiniteFlow for the reduction rule associated to the integral. 

For example, in sector $\vec m=(1,1,1,0,0)$, we still have the integrals
\begin{equation}
    F(1,1,1,0,-1), \quad F(1,1,1,n_{4},0)\, .
\end{equation}
For the fixed integral $F(1,1,1,0,-1)$, we take our identities and input seeds up to rank $1$
\begin{equation}
    \{(1,1,1,0,0),(1,1,1,-1,0),(1,1,1,0,-1)\}\,,
\end{equation}
and attempt to solve these equations for this integral. We find no reduction rules that reduce it to lower-weight integrals, so we interpret it as a master integral.

For the integral $F(1,1,1,n_{4},0)$, we take our identities and input seeds of increasing rank one at a time
\begin{equation}\label{eq:seeds}
    \{(1,1,1,n_{4},0),(1,1,1,n_{4}-1,0),(1,1,1,n_{4},-1),(1,1,1,n_{4}-1,-1),\dots\}\,,
\end{equation}
each time we attempt to solve the system of equations for reduction rules that would allow us to reduce integrals of this type. In this case, we find the rule after inputting the first $2$ seeds in Eq. \ref{eq:seeds}
\begin{equation}
    F(1,1,1,n_{4}-2,0)\rightarrow \text{lower-weight integrals}\,.
\end{equation}
This allows us to reduce all integrals of this type except for $F(1,1,1,0,0)$, $F(1,1,1,-1,0)$, which we can find no further reduction rules for. We also find that the denominator on the right hand side vanishes for $n_{4}=0$, as shown in Appendix \ref{app:Example}, and therefore we also cannot reduce $F(1,1,1,-2,0)$. This is the final piece, and with this we find that all integrals in this sector can be reduced by $3$ reduction rules, except for the master integrals
\begin{equation}
    F(1,1,1,0,0), \quad F(1,1,1,-1,0), \quad F(1,1,1,0,-1), \quad F(1,1,1,-2,0)\,.
\end{equation}
These are the master integrals if one does not include symmetry
relations. If symmetry relations are included, there are in fact only
$2$ master integrals. In appendix $\ref{app:Example}$, we provide a
table describing which integrals can be solved this way and what input
seeds are required. While the neighbourhood size in this step is very
heuristic, in practice a small amount of neighbouring seed integrals
are usually needed, and we have always found the correct number of
master integrals.

\paragraph{Summary} 

\begin{figure}[h!]
    \centering
    \includegraphics[scale=0.8]{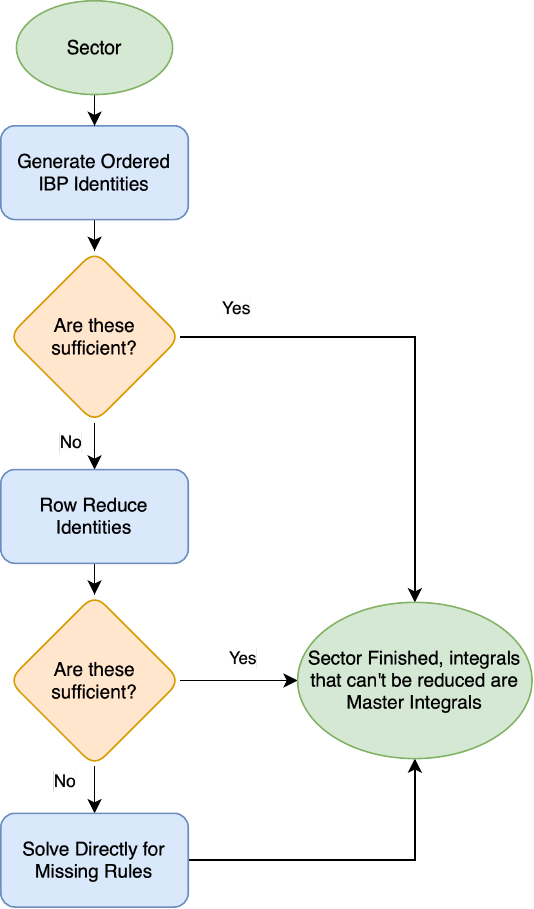}
    \caption{Flow Chart to represent the first stage of the algorithm, namely building reduction rules.\label{fig:flowchart}}
\end{figure}

To summarize this stage of the algorithm, we have reshuffled our initial syzygy-generated identities on a given sector in order to find a minimal set of reduction rules that will reduce any integral in this sector, except for the master integrals. We represent how this first stage of the algorithm works with the flow chart in Fig. \ref{fig:flowchart}

\subsection{Applying Reduction Rules}

The second stage of this algorithm is to apply the reduction rules built in the first stage to fully reduce a set of target integrals.

One way to complete this step is to iteratively apply the reduction rules to our target integrals, obtaining more lower weight integrals each time. After continuous application this number of integrals would start to collapse down and eventually the expressions will reduce to the master integrals. One efficient way of performing this step would be to formulate this iterative application of rules as a string of matrices to be applied to the master integrals. For example, if you start with a list of target integrals $F(\vec n_{i})$, $i=1,\dots,T_{0}$, the application of the first reduction rule will look like
\begin{equation}
    \begin{pmatrix}
        F(\vec n_{1})\\
        F(\vec n_{2})\\
        \vdots\\
        F(\vec n_{T_{0}})
    \end{pmatrix} = \begin{pmatrix}
        &&\\
        &T_{0}\times T_{1}&\\
        &&
    \end{pmatrix}\begin{pmatrix}
        F(\vec n_{1}')\\
        F(\vec n_{2}')\\
        \vdots\\
        F(\vec n_{T_{1}}')
    \end{pmatrix}\,.
\end{equation}
Here, the $T_{0}\times T_{1}$ matrix will be given by the reduction rules, and the $T_{1}$ new integrals are of lower weight than the first $T_{0}$ target integrals. We can then apply the same procedure to the $T_{1}$ new integrals, which we would now consider to be the target integrals. Performing this step iteratively, we will eventually be left with only the master integrals $J_{i}$ as our target, and a string of matrices
\begin{equation}
    \begin{pmatrix}
        F(\vec n_{1})\\
        F(\vec n_{2})\\
        \vdots\\
        F(\vec n_{T_{0}})
    \end{pmatrix} = \begin{pmatrix}
        &&\\
        &T_{0}\times T_{1}&\\
        &&
    \end{pmatrix}\begin{pmatrix}
        &&\\
        &T_{1}\times T_{2}&\\
        &&
    \end{pmatrix}\dots\begin{pmatrix}
        &&\\
        &T_{k}\times \nu&\\
        &&
    \end{pmatrix}\begin{pmatrix}
        J_{1}\\
        J_{2}\\
        \vdots\\
        J_{\nu}
    \end{pmatrix}\,.
\end{equation}
Then, to find the reduction we just need to perform a matrix multiplication of this string from left to right, which can be done using a finite field reconstruction algorithm to recover analytic dependence on parameters.

\begin{figure}[h!]
    \centering
    \includegraphics[scale=0.8]{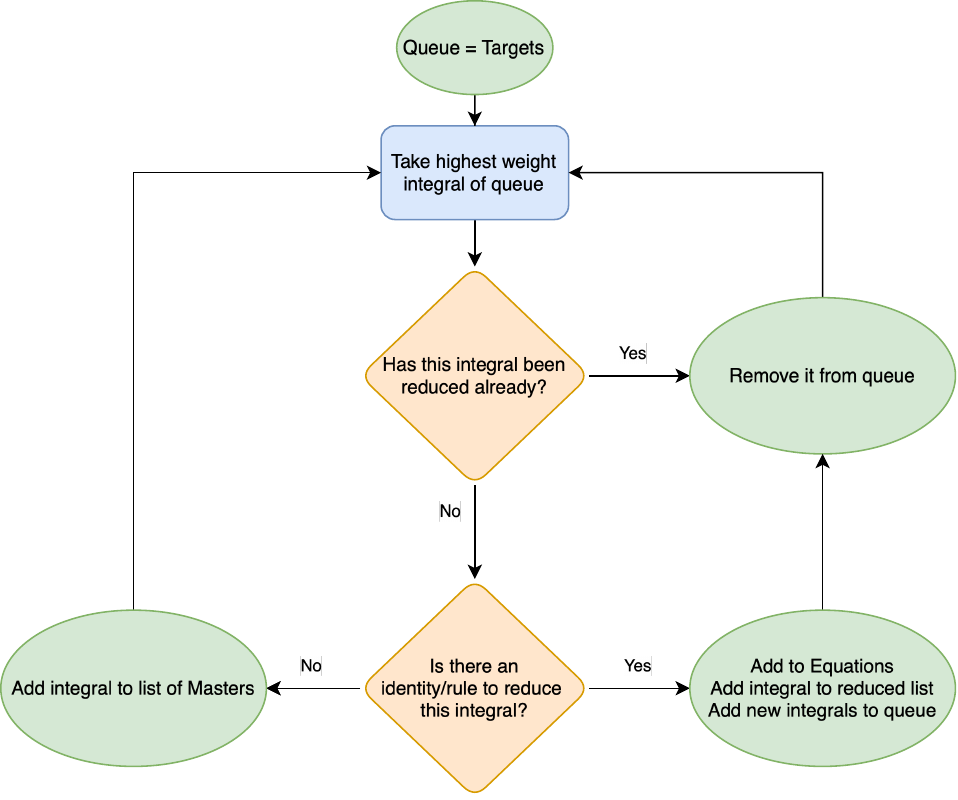}
    \caption{Flow Chart to represent the second stage of the algorithm, namely applying reduction rules.\label{fig:equations}}
\end{figure}

The other way to complete this step, is to use these reduction rules to generate a full system of equations with distinct highest-weight integrals and perform backward substitution. We do this by starting with a list of target integrals, which we call the \textit{queue}. We can then cycle through all integrals in the queue one by one and for each one generate an equation for which this integral is the highest weight one within that equation, so the equation will act as a reduction rule to reduce this integral to lower weight ones. We then add the appropriate equation to our list of equations and add any new integrals that this equation introduces to the queue, as well as add the reduced integral to a list of reduced integrals. If there doesn't exist an identity or rule to reduce an integral in the queue it is added to the list of master integrals. This is described in the flowchart in Fig. \ref{fig:equations}

Doing this iteratively until the queue is empty, we are left with a set of equations, a set of reduced integrals and a set of master integrals. We can then solve this system to find the reduction of the target integrals into the master integrals. Solving a linear system using Gaussian elimination comes in two steps; the first step is the forward elimination, which eliminates variables and puts the matrix into a row echelon form (or an upper triangular form in the special case of a full-rank square matrix), and the second step is the backward substitution, which solves for the unknowns starting from the last row of the matrix and brings the matrix to a \emph{reduced} row echelon form. This is depicted below for a simple example below, where the last column of the matrix corresponds to the only "master integral",
\begin{equation}\label{eq:backsub}
    \begin{pmatrix}
        \#&\#&\#&0\\
        \#&\#&0&\#\\
        \#&0&0&\#\\
        0&\#&\#&\#\\
        0&0&\#&0
    \end{pmatrix}\xrightarrow{\text{forward elimination}}\begin{pmatrix}
        \#&\#&\#&0\\
        0&\#&\#&\#\\
        0&0&\#&\#\\
        0&0&0&\#\\
        0&0&0&0
    \end{pmatrix}\xrightarrow{\text{backward substitution}}\begin{pmatrix}
        \#&0&0&\#\\
        0&\#&0&\#\\
        0&0&\#&\#\\
        0&0&0&\#\\
        0&0&0&0
    \end{pmatrix}\,.
\end{equation}

In the context of IBP reduction, forward elimination essentially turns the system of equations into reduction rules that express each non-master integral in terms of lower-weight integrals that are generally not master integrals, while backward substitution finishes the job of expressing each non-master integral in terms of linear combinations of master integrals.
Since our system of equations will naturally be in a row echelon form (the second matrix in Eq. \ref{eq:backsub}), as they are generated using reduction rules, the only step required is the backward substitution, which is typically a far faster step than forward elimination for numerical (finite-field) Gaussian elimination.

For the sunrise example, we generate a system of 690 equations and 697 variables, that can reduce these target integrals to 7 master integrals
\begin{equation}
    \left\{
    \begin{aligned}
    &F(1,1,1,0,0),F(1,1,1,-1,0),F(1,1,1,0,-1),F(1,1,1,-2,0)\\
    &\quad \qquad \qquad F(1,1,0,0,0),F(1,0,1,0,0),F(0,1,1,0,0)
    \end{aligned}
    \right\}\,,
\end{equation}
and in the process we are able to reduce 690 integrals. Further details are in Appendix \ref{app:Example}.

\section{Examples \label{sec:ex}}

In this section we present some examples of complex integrals belonging to non-trivial families that we have managed to reduce using this method. The implementation of our algorithm was written in Mathematica and used FiniteFlow \cite{Peraro:2019svx} and Singular \cite{DGPS}.

\subsection{Double Box with External Mass}

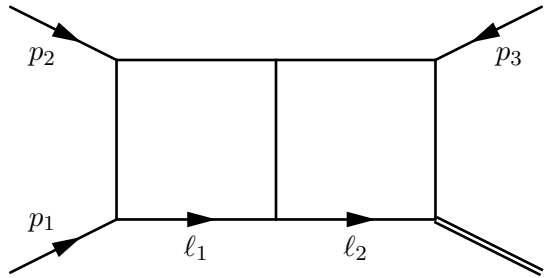
\begin{figure}[h!]
    \centering
    \begin{fmfgraph*}(200,100)
        \fmfforce{(0,0)}{i1}
        \fmfforce{(0,h)}{i2}
        \fmfforce{(w,0)}{i3}
        \fmfforce{(w,h)}{i4}
        \fmfforce{(0.2w,0.2h)}{v1}
        \fmfforce{(0.2w,0.8h)}{v2}
        \fmfforce{(0.5w,0.2h)}{v3}
        \fmfforce{(0.5w,0.8h)}{v4}
        \fmfforce{(0.8w,0.2h)}{v5}
        \fmfforce{(0.8w,0.8h)}{v6}
        \fmf{fermion,label=$p_{1}$}{i1,v1}
        \fmf{fermion,label=$p_{2}$}{i2,v2}
        \fmf{fermion,label=$p_{3}$}{i4,v6}
        \fmf{dbl_plain}{i3,v5}
        \fmf{fermion,label=$\ell_{1}$}{v1,v3}
        \fmf{fermion,label=$\ell_{2}$}{v3,v5}
        \fmf{plain}{v2,v4,v6}
        \fmf{plain}{v1,v2}
        \fmf{plain}{v3,v4}
        \fmf{plain}{v5,v6}
    \end{fmfgraph*}
    \caption{Two Loop Box\label{fig:doublebox}}
\end{figure}

The first example is that of the double box with an external mass, as shown in Fig. \ref{fig:doublebox}

This has propagators
\begin{equation}
    \begin{aligned}
    \rho_{1} &= \ell_{1}^{2}, \quad \rho_{2} = (\ell_{1}-p_{1})^{2}, \quad \rho_{3} = (\ell_{1}-p_{1}-p_{2})^{2}\\
    \rho_{4} &= (\ell_{2}-p_{1}-p_{2})^{2}, \quad \rho_{5} = (\ell_{2}-p_{1}-p_{2}-p_{3})^{2}, \quad \rho_{6} = \ell_{2}^{2}\\
    \rho_{7} &= (\ell_{2}-\ell_{1})^{2}, \quad \rho_{8} = (\ell_{1}-p_{1}-p_{2}-p_{3})^{2}, \quad \rho_{9} = (\ell_{2}-p_{1})^{2}\,.
    \end{aligned}
\end{equation}
Here, $\rho_{1\dots7}$ are denominators and $\rho_{8}$ and $\rho_{9}$ are numerators. The kinematics are
\begin{equation}
    p_{i}^{2}=0, \quad s = (p_{1}+p_{2})^{2}, \quad t = (p_{2}+p_{3})^{2}, \quad u = (p_{1}+p_{3})^{2} = m^{2}-s-t\,.
\end{equation}
We attempt to reduce the integrals
\begin{equation}
    \{F(1,1,1,1,1,1,1,-10,-10),F(1,2,1,1,1,1,1,-6,-6),F(1,1,1,1,1,1,1,-2,-15)\}\,,
\end{equation}
which go up to rank $20$ with one dot. 
\begin{table}
\centering
    \begin{tabular}{|c|c|c|c|c|}
        \hline
        Cut&Time Taken&Number of Equations&Number of Masters\\
        \hline
        \{5,7\}&284s&18971&14\\
        \hline
        \{1,4,7\}&60s&8120&4\\
        \hline
        \{3,6,7\}&178s&12287&8\\
        \hline
        \{4,6,7\}&57s&2643&4\\
        \hline
        \{1,3,4,6\}&26s&1153&3\\
        \hline
        \{1,3,5,6\}&35s&2031&5\\
        \hline
    \end{tabular}
    \caption{Table showing the time taken for generating the system of reduction rules \textit{analytically} on a set of spanning cuts, as well as the number of rules applied (equations) and master integrals.\label{tab:tab2}}
\end{table}
We provide statistics for the reduction of this set of integrals in Table~\ref{tab:tab2}. The
  sample time for applying these reduction rules using backward
  substitution at a single numerical point (in the space of kinematic
  and spacetime dimension parameters) modulo some prime number is very
  small, however it is well known that the IBP coefficients for the
  reduction of these high-rank integrals are highly non-trivial, and
  therefore the rational function reconstruction would still be a
  bottleneck here, however for numerical reduction this is not a
  problem.

In general, the number of equations generated in this case is much smaller than what would be needed for the Laporta system. This is not a direct comparison as the Laporta system would consist of a large number of sparse equations, and the resulting system we have obtained here will be similar to the Laporta system after a very time-consuming step of forward-elimination. However this forward-elimination step is done at the level of IBP operators, so we benefit from not having to work with this large sparse system of equations. We have included, as an ancillary file, example reduction rule identities for cut $\{1,4,7\}$ and a (very) simple implementation of them.

\subsection{Massless Pentabox}

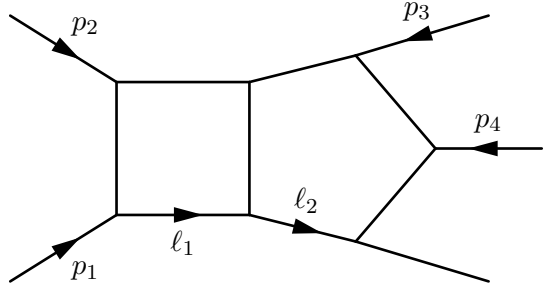
\begin{figure}[h!]
    \centering
    \begin{fmfgraph*}(200,100)
        \fmfforce{(0,0)}{i1}
        \fmfforce{(0,h)}{i2}
        \fmfforce{(0.9w,0)}{i3}
        \fmfforce{(0.9w,h)}{i4}
        \fmfforce{(w,0.5h)}{i5}
        \fmfforce{(0.2w,0.25h)}{v1}
        \fmfforce{(0.2w,0.75h)}{v2}
        \fmfforce{(0.45w,0.25h)}{v3}
        \fmfforce{(0.45w,0.75h)}{v4}
        \fmfforce{(0.65w,0.15h)}{v5}
        \fmfforce{(0.65w,0.85h)}{v6}
        \fmfforce{(0.8w,0.5h)}{v7}
        \fmf{fermion,label=$p_{1}$}{i1,v1}
        \fmf{fermion,label=$p_{2}$}{i2,v2}
        \fmf{fermion,label=$p_{3}$}{i4,v6}
        \fmf{fermion,label=$p_{4}$}{i5,v7}
        \fmf{plain}{i3,v5}
        \fmf{fermion,label=$\ell_{1}$}{v1,v3}
        \fmf{fermion,label=$\ell_{2}$}{v3,v5}
        \fmf{plain}{v2,v4,v6,v7,v5}
        \fmf{plain}{v1,v2}
        \fmf{plain}{v3,v4}
    \end{fmfgraph*}
    \caption{Massless Pentabox\label{fig:pentabox}}
\end{figure}

The second example we present is the massless pentabox, as shown in Fig. \ref{fig:pentabox}

In this case, the propagators are 
\begin{equation}
    \begin{aligned}
    \rho_{1} &= \ell_{1}^{2}, \quad \rho_{2} = (\ell_{1}-p_{1})^{2}, \quad \rho_{3} = (\ell_{1}-p_{1}-p_{2})^{2}, \quad \rho_{4} = (\ell_{2}-p_{1}-p_{2})^{2}\\ 
    \rho_{5} &= (\ell_{2}-p_{1}-p_{2}-p_{3})^{2}, \quad \rho_{6} = (\ell_{2}-p_{1}-p_{2}-p_{3}-p_{4})^{2}, 
    \quad \rho_{7} = \ell_{2}^{2}\\
    \rho_{8} &= (\ell_{2}-\ell_{1})^{2}, \quad \rho_{9} = (\ell_{1}-p_{1}-p_{2}-p_{3})^{2}, \quad \rho_{10} = (\ell_{1}-p_{1}-p_{2}-p_{3}-p_{4})^{2}\\
    \rho_{11} &= (\ell_{2}-p_{1})^{2}\,,
    \end{aligned}
\end{equation}
where $\rho_{1\dots8}$ are denominators and $\rho_{9}$, $\rho_{10}$ and $\rho_{11}$ are numerators. The kinematics are
\begin{equation}
    p_{i}^{2} = 0, \quad s_{ij} = (p_{i}+p_{j})^{2}, \quad s_{34} = -s_{12}-s_{13}-s_{14}-s_{23}-s_{24}\,.
\end{equation}
and due to the complexity we choose to work at the numerical point
\begin{equation}
    \{s_{12}\rightarrow1, s_{13}\rightarrow-3, s_{14}\rightarrow7, s_{23}\rightarrow-11, s_{24} \rightarrow13\}\,.
\end{equation}
We attempt to reduce the random set of integrals
\begin{equation}
    \{F(1,1,1,1,1,1,1,1,-10,-10,0),F(1,1,1,1,1,1,1,1,-5,-6,-3)\}\,,
\end{equation}
which go up to rank $20$. 
\begin{table}
\centering
    \begin{tabular}{|c|c|c|c|c|}
        \hline
        Cut&Time Taken&Number of Equations&Number of Masters\\
        \hline
        \{1,4,8\}&6997s&51619&21\\
        \hline
        \{1,5,8\}&2912s&39446&27\\
        \hline
        \{2,5,8\}&23721s&112188&31\\
        \hline
        \{1,3,4,6\}&974s&3979&13\\
        \hline
        \{1,3,4,7\}&1063s&4275&9\\
        \hline
        \{2,4,7,8\}&7064s&28338&12\\
        \hline
    \end{tabular}
    \caption{Table showing the time taken for generating the system of reduction rules at the given numerical point on a set of spanning cuts (cuts that can be obtained from these ones through symmetry relations are omitted), as well as the number of rules applied (equations) and master integrals.\label{tab:tab3}}
\end{table}  
We provide statistics for the reduction of this set of integrals in Table~\ref{tab:tab3}. We see here that in some cases, in particular cut $\{2,5,8\}$, the system takes a significant amount of time to generate, however the respective Laporta system is so large in this case that it is prohibitive to solve. For example, we attempted to reduce the less challenging rank-$14$ integral $F(1,1,1,1,1,1,1,1,-5,-6,-3)$ with Kira 3 \cite{Lange:2025fba}, and the job was killed after 23 minutes for running out of RAM whilst generating equations. (The laptop had 64 GB of total RAM, and 55.7 GB was consumed by Kira 3 when the job was killed.)

\subsection{Spinning Black Hole}

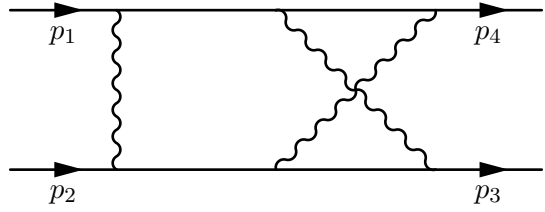
\begin{figure}[h!]
    \centering
    \begin{fmfgraph*}(200,100)
        \fmfforce{(0,0.2h)}{i1}
        \fmfforce{(0,0.8h)}{i2}
        \fmfforce{(w,0.2h)}{i3}
        \fmfforce{(w,0.8h)}{i4}
        \fmfforce{(0.2w,0.2h)}{v1}
        \fmfforce{(0.2w,0.8h)}{v2}
        \fmfforce{(0.5w,0.2h)}{v3}
        \fmfforce{(0.5w,0.8h)}{v4}
        \fmfforce{(0.8w,0.2h)}{v5}
        \fmfforce{(0.8w,0.8h)}{v6}
        \fmf{fermion,label=$p_{2}$}{i1,v1}
        \fmf{fermion,label=$p_{1}$}{i2,v2}
        \fmf{fermion,label=$p_{4}$}{v6,i4}
        \fmf{fermion,label=$p_{3}$}{v5,i3}
        \fmf{plain}{v1,v3,v5}
        \fmf{plain}{v2,v4,v6}
        \fmf{wiggly}{v1,v2}
        \fmf{wiggly}{v3,v6}
        \fmf{wiggly}{v5,v4}
    \end{fmfgraph*}
    \caption{Post-Minkowskian Non-Planar Double Box\label{fig:npdoublebox}}
\end{figure}

To illustrate the particular relevance of this paper, we also provide a physical example. We refer directly to \cite{Akpinar:2025bkt}, which provided the calculation of the quartic-in-spin, third post-Minkowskian order correction to the binary system of a spinning and a spinless black hole. The most troublesome part of this calculation was the IBP reduction, for which the most complex diagram was the nonplanar double box, shown in Fig. \ref{fig:npdoublebox}

We use the same conventions as this paper, and \cite{Parra-Martinez:2020dzs}. The external kinematics can be parametrized by
\begin{equation}
    \begin{aligned}
    p_{1} &= -\bar{p}_{1}+q/2, \quad p_{4} = \bar{p}_{1}+q/2\,,\\
    p_{2} &= -\bar{p}_{2}-q/2, \quad p_{3} = \bar{p}_{2}-q/2\,,
    \end{aligned}
\end{equation}
where $q = p_{1}+p_{4}$, and $p_{1}^{2}=p_{4}^{2}=m_{1}^{2}$, $p_{2}^{2}=p_{3}^{2}=m_{2}^{2}$. We then define
\begin{equation}
    u_{i} = \frac{\bar{p}_{i}}{\bar{m}_{i}}, \quad \bar{m}_{i} = m_{i}^{2}-q^{2}/4\,,
\end{equation}
such that
\begin{equation}
    u_{i}\cdot q=0\,.
\end{equation}
With these conventions, we find that the matter propagators become, in the classical limit $|q| \ll |p_i|$,
\begin{equation}
    \frac{1}{(\ell+\bar{p}_{i}\pm q/2)^{2}-m_{i}^{2}} = \frac{1}{2\bar{p}_{i}\cdot \ell}-\frac{\ell^{2}\pm\ell\cdot q}{(2\bar{p}_{i}\cdot\ell)^{2}}+\dots\,.
\end{equation}
Therefore we can associate linearized propagators of the form $u_{i}\cdot\ell_{a}$ to our matter propagators. The list of propagators that defines this family is
\begin{equation}
    \begin{aligned}
        &\rho_{1} = 2u_{1}\cdot\ell_{1}, \quad \rho_{2} = -2u_{2}\cdot\ell_{1}, \quad \rho_{3} = 2u_{1}\cdot\ell_{2}, \quad \rho_{4} = -2u_{2}\cdot(\ell_{1}-\ell_{2})\\
        \rho_{5} &= \ell_{1}^{2}, \quad \rho_{6} = \ell_{2}^{2}, \quad \rho_{7} = (\ell_{1}-\ell_{2}-q)^{2}, \quad \rho_{8} = (\ell_{1}-q)^{2}, \quad \rho_{9} = (\ell_{2}+q)^{2}\,,
    \end{aligned}
\end{equation}
and for the kinematics we define the dimensionless variable $y=u_{1}\cdot u_{2}$ and set the only dimensionful parameter $q^{2}=-1$. 

Here, $\rho_{1,\dots,4}$ represent the matter propagators, and $\rho_{5,\dots,7}$ represent the graviton propagators. All graviton propagators must be present in an integral for it to contribute in the classical limit, as otherwise there would be a contact term between the two black holes. Furthermore, these graviton propagators only appear as single powers in the denominator, so we are free to perform the reduction on the triple cut of these $3$ propagators. 

In this particular problem, there are $47365$ target integrals to reduce, and these can be up to rank $10$ with $10$ dots. For example
\begin{equation}
    F(11,1,1,1,1,1,1,-10,0),F(3,5,1,5,1,1,1,-7,-3),\dots\,,
\end{equation}
are integrals we need to reduce. In \cite{Akpinar:2025bkt}, the IBP reduction was completed for these integrals using a private version of FIRE which builds upon Refs.\ \cite{Smirnov:2019qkx,Smirnov:2023yhb,Smirnov:2024onl} at $\sim1000$ numerical points of $y$ and $\D$. Each numerical point took $\sim15$ minutes to evaluate, therefore the total run time ended up being $\sim10$ days on a cluster.

We attempted to perform this IBP reduction on a laptop using the methods discussed in this paper. Whereas the equation generation for FIRE was very fast, it took $\sim45$ minutes with our algorithm to build the reduction rules, and then $\sim8$ hours to generate the equations. This length of time is primarily due to the large number of target integrals entering the queue at the beginning of the algorithm. The system of equations was built with analytic dependence on $y$ and $\D$, and it contained $152113$ equations and $152104$ variables. However the slower equation generation is more than compensated by the faster solving time. We are able to solve these equations at different numerical points modulo a $32$-bit prime, with each numerical point taking $\sim8s$ to evaluate, therefore the solving time would amount to $\sim2$ hours. Combined, this gives a total time of $\sim11$ hours.

\section{Conclusion \label{sec:conclusion}}

In this paper, we presented a novel algorithm for reducing Feynman integrals by generating symbolic reduction rules that can be applied to an arbitrary set of target integrals. This method is based on generating syzygy-constrained IBP identities without raised propagator powers sector-by-sector and reshuffling these identities to obtain a complete set of reduction rules. This reshuffling essentially amounts to performing forward elimination at the level of IBP operators, and if necessary, solving a small linear system of IBP equations obtained from a small set of seed integrals in the neighborhood of the target integral. We can then apply these reduction rules to a set of target integrals either iteratively, or via seeding and backward substitution.

The motivation behind this algorithm is for the reduction of Feynman integrals with high powers of numerators or propagators, a demanding computational step in evaluating multi-loop scattering amplitudes. We therefore tested this algorithm against two highly non-trivial examples of rank-$20$ integrals for the double box with an external mass and the massless pentabox. In both these cases we found we were able to improve significantly the solving step of the calculation, compared with the state of the art---for example, even rank-$14$ pentabox integrals are beyond the reach of Kira 3 on a laptop with 64 GB of RAM, since the system of equations becomes too large. We also presented an application of this algorithm to a physical problem, namely the IBP reduction related to the scattering amplitude for a spinning black hole binary system at third post-Minkowskian order. This application perfectly illustrates the importance of this new algorithm, as it is clear that these complex integrals do appear in real calculations. We found that in this example, the algorithm presented here reduced the overall run time from $\sim10$ days to $\sim11$ hours.

The ideas presented in this work open many possibilities for further
exploration. First of all, this algorithm can be incredibly effective
for the computation of amplitudes in non-renormalizable field theories
such as gravity, where it is common to find very complex integrals
that need to be reduced. In this paper we showed this was the case for
a spinning black hole calculation \cite{Akpinar:2025bkt}, and we
expect that there are many other calculations that could benefit from
applying these techniques. Second, other variants of our method, e.g.\
based on IBP identities in the original Laporta approach without
additional constraints from syzygy equations, could prove useful for
complicated diagram topologies for which it is difficult to solve the
syzygy equations. Indeed, we have found that in some cases, the use of
syzygy constraints does not have a significant impact on performance
time, however, there are also many cases where a similar approach with
standard Laporta identities is significantly slower or unable to
finish. Third, it would prove interesting to study further the
technical aspects of this algorithm, such as the monomial ordering
chosen, and whether there is some kind of non-commutative algebra
structure present in the generated reduction rules, which has been
found for IBP operators in the past (see e.g.\ \cite{Barakat:2022qlc,Barakat:2022ttc}). Finally, our results indicate that despite decades of work, significant improvements to the computational efficiency of IBP reduction are still possible, and intense future investigations are warranted.

\paragraph{Note added:} During the final stage of preparing this manuscript, we learnt about an updated version of the work \cite{Brunello:2024tqf}, in which the reduction of pentabox integrals up to rank $20$ was also achieved through intersection theory \cite{Mastrolia:2018uzb,Frellesvig:2019kgj,Frellesvig:2019uqt,Frellesvig:2020qot,Chestnov:2022xsy,Brunello:2023rpq}. We have used our method to perform a reduction of the same integrals presented in this work, and have found agreement.

\section*{Acknowledgements}

We wish to thank William J. Torres Bobadilla for organising the \textit{Scattering Amplitudes @ Liverpool} workshop earlier this year, and the participants for many interesting discussions related to this work during the workshop. We also wish to thank Giulio Crisanti for interesting discussions related to this work. S.S. thanks the Higgs Centre for Theoretical Physics at the University of Edinburgh for hosting him during the development of this project. S.S. research is partially supported by the Amplitudes INFN scientific initiative.
M.Z.’s work is supported in part by the U.K.\ Royal Society through Grant URF\textbackslash R1\textbackslash 20109. For the purpose of open access, the authors have applied a Creative Commons Attribution (CC BY) license to any Author Accepted Manuscript version arising from this submission.

\newpage

\appendix
\section{Example: Equal Mass Sunrise\label{app:Example}}

In this appendix we provide more details for the equal mass sunrise discussed in Section \ref{sec:algorithm}.

\paragraph{Generating Ordered Identities}

We provide the Module matrix for the sectors $(2,1,1,0,0)$ and $(1,1,1,0,0)$ \footnotesize
\begin{equation*}
    M = \left(
\begin{array}{ccc}
 2 m^2+2 \rho_{1} & 0 & m^2+\rho_{1}-\rho_{2}+\rho_{3}+2 \rho_{5}-s \\
 -m^2-\rho_{1}-\rho_{2}+\rho_{3}+2 \rho_{4}+2 \rho_{5}-s & 0 & m^2-\rho_{1}+\rho_{2}+\rho_{3}+2 \rho_{4}-s \\
 2 \rho_{4} & 0 & 2 \rho_{4}+2 \rho_{5}-2 s \\
 0 & -m^2-\rho_{1}-\rho_{2}+\rho_{3}+2 \rho_{4}+2 \rho_{5}-s & m^2+\rho_{1}-\rho_{2}+\rho_{3}+2 \rho_{5}-s \\
 0 & 2 m^2+2 \rho_{2} & m^2-\rho_{1}+\rho_{2}+\rho_{3}+2 \rho_{4}-s \\
 0 & 2 \rho_{5} & 2 \rho_{4}+2 \rho_{5}-2 s \\
 -\rho_{1} & 0 & 0 \\
 0 & -\rho_{2} & 0 \\
 0 & 0 & -\rho_{3} \\
\end{array}
\right)
\end{equation*}
\normalsize
for this there are $36$ syzygy solutions $\vec c^{T}M=0$.The Module matrix for the sectors $(2,1,0,0,0)$ and $(1,1,0,0,0)$ is
\footnotesize
\begin{equation*}
    M = \left(
\begin{array}{cc}
 2 m^2+2 \rho_{1} & 0 \\
 -m^2-\rho_{1}-\rho_{2}+\rho_{3}+2 \rho_{4}+2 \rho_{5}-s & 0 \\
 2 \rho_{4} & 0 \\
 0 & -m^2-\rho_{1}-\rho_{2}+\rho_{3}+2 \rho_{4}+2 \rho_{5}-s \\
 0 & 2 m^2+2 \rho_{2} \\
 0 & 2 \rho_{5} \\
 -\rho_{1} & 0 \\
 0 & -\rho_{2} \\
\end{array}
\right)
\end{equation*}
\normalsize
for this there are $12$ syzygy solutions $\vec c^{T}M=0$.

For sector $(2,1,1,0,0)$, the relevant identities have the following leading weight terms
\begin{equation}
    \begin{aligned}
    (4n_{4}-2n_{2})F(2,1,1,n_{4},n_{5}-1)+\dots&=0\\
    -2F(2,1,1,n_{4},n_{5}-2)+\dots&=0\\
    12F(2,1,1,n_{4},n_{5}-2)+\dots&=0\\
    -4F(2,1,1,n_{4},n_{5}-2)+\dots&=0\\
    6s(2n_{4}-n_{5})F(2,1,1,n_{4},n_{5}-1)+\dots&=0\\
    -2F(2,1,1,n_{4},n_{5}-2)+\dots&=0\\
    -4F(2,1,1,n_{4}-1,n_{5}-1)+\dots&=0\\
    -4F(2,1,1,n_{4}-1,n_{5}-2)+\dots&=0\\
    4(n_{5}-\D)F(2,1,1,n_{4},n_{5}-3)+\dots&=0\\
    12(\D-6+n_{5}-4n_{4})F(2,1,1,n_{4},n_{5}-2)+\dots&=0\\
    4(2-\D+n_{5})F(2,1,1,n_{4},n_{5}-2)+\dots&=0\\
    2F(2,1,1,n_{4},n_{5}-2)+\dots&=0\\
    2(2-\D+2n_{4})F(2,1,1,n_{4}-1,n_{5}-2)+\dots&=0
    \end{aligned}
\end{equation}
For sector $(1,1,1,0,0)$, the relevant identities are
\begin{equation}
    \begin{aligned}
    2(2n_{4}-n_{5})F(1,1,1,n_{4},n_{5}-1)+\dots&=0\\
    6s(2n_{4}-n_{5})F(1,1,1,n_{4},n_{5}-1)+\dots&=0\\
    4(n_{5}-\D)F(1,1,1,n_{4},n_{5}-3)+\dots&=0\\
    12(\D-3+n_{5}-4n_{4})F(1,1,1,n_{4},n_{5}-2)+\dots&=0\\
    4(1-\D+n_{5})F(1,1,1,n_{4},n_{5}-2)+\dots&=0\\
    2(2n_{4}-\D)F(1,1,1,n_{4}-1,n_{5}-2)+\dots&=0
    \end{aligned}
\end{equation}
The first of these identities we give fully, for the interested reader
\begin{equation}\label{eq:fullId}
    \begin{aligned}
        0&=2 (2 n_{4}-n_{5}) F(1,1,1,n_{4},n_{5}-1)-2 n_{4} \left(m^2+s\right) F(1,1,1,n_{4}+1,n_{5}-1)\\
        &+n_{4} s \left(m^2+s\right) F(1,1,1,n_{4}+1,n_{5})-(n_{4}-n_{5}) \left(m^2+3 s\right) F(1,1,1,n_{4},n_{5})\\
        &+2 n_{5} \left(m^2+s\right) F(1,1,1,n_{4}-1,n_{5}+1)-n_{5} s \left(m^2+s\right) F(1,1,1,n_{4},n_{5}+1)\\
        &+n_{4} s F(0,1,1,n_{4}+1,n_{5})+n_{4} s F(1,0,1,n_{4}+1,n_{5})-n_{4} s F(1,1,0,n_{4}+1,n_{5})\\
        &-n_{5} s F(0,1,1,n_{4},n_{5}+1)-n_{5} s F(1,0,1,n_{4},n_{5}+1)+n_{5} s F(1,1,0,n_{4},n_{5}+1)\\
        &-2 n_{4} F(0,1,1,n_{4}+1,n_{5}-1)+(n_{4}-n_{5}) F(1,1,0,n_{4},n_{5})\\
        &+2 (n_{4}-2 n_{5}) F(1,1,1,n_{4}-1,n_{5})+2 n_{5} F(1,0,1,n_{4}-1,n_{5}+1)\\
        &+(n_{5}-n_{4}) F(0,1,1,n_{4},n_{5})+(n_{5}-n_{4}) F(1,0,1,n_{4},n_{5})\,.
    \end{aligned}
\end{equation}
For sector $(2,1,0,0,0)$, the relevant identities are
\begin{equation}
    \begin{aligned}
        -2F(2,1,n_{3},n_{4}-1,n_{5})+\dots&=0\\
        2(n_{3}-n_{4})F(2,1,n_{3},n_{4},n_{5}-1)+\dots&=0\\
        (\D-1-n_{3}-n_{5})F(2,1,n_{3},n_{4},n_{5}-1)+\dots&=0\\
        2F(2,1,n_{3}-1,n_{4},n_{5})+\dots&=0\\
        (3-\D+n_{3}+n_{4})F(2,1,n_{3}-1,n_{4},n_{5})+\dots&=0\\
        (1-\D+n_{3}+n_{5})F(2,1,n_{3}-1,n_{4},n_{5})+\dots&=0\\
        (1-\D+n_{3}+n_{5})F(2,1,n_{3},n_{4},n_{5}-1)+\dots&=0\\
        -2m^{2}F(2,1,n_{3},n_{4},n_{5})+\dots&=0
    \end{aligned}
\end{equation}
For sector $(1,1,0,0,0)$, the relevant identities are
\begin{equation}
    \begin{aligned}
        2(n_{3}-n_{4})F(1,1,n_{3},n_{4},n_{5}-1)+\dots&=0\\
        (\D-1-n_{3}-n_{5})F(1,1,n_{3},n_{4},n_{5}-1)+\dots&=0\\
        (1-\D+n_{3}+n_{4})F(1,1,n_{3}-1,n_{4},n_{5})+\dots&=0\\
        (1-\D+n_{3}+n_{4})F(1,1,n_{3},n_{4}-1,n_{5})+\dots&=0\\
        (1-\D+n_{3}+n_{5})F(1,1,n_{3}-1,n_{4},n_{5})+\dots&=0\\
        (1-\D+n_{3}+n_{5})F(1,1,n_{3},n_{4},n_{5}-1)+\dots&=0\\
    \end{aligned}
\end{equation}

The used identities are as follows:
\begin{center}
    \begin{tabular}{|c|c|c|}
        \hline
        Sector&Used Identities (leading weight terms)&Irreducible Integrals\\
        \hline
        $(2,1,1,0,0)$&\makecell{$-2F(2,1,1,n_{4},n_{5}-2)$\\ $-4F(2,1,1,n_{4}-1,n_{5}-1)$}&\makecell{$F(2,1,1,n_{4},0)$\\ $F(2,1,1,0,-1)$}\\
        \hline
        $(1,1,1,0,0)$&\makecell{$4(1-\D+n_{5})F(1,1,1,n_{4},n_{5}-2)$\\$2(2n_{4}-n_{5})F(1,1,1,n_{4},n_{5}-1)$}&\makecell{$F(1,1,1,0,-1)$\\ $F(1,1,1,n_{4},0)$}\\
        \hline
        $(2,1,0,0,0)$&$-2m^{2}F(2,1,n_{3},n_{4},n_{5})$&N/A\\
        \hline
        $(1,1,0,0,0)$&\makecell{$(1-\D+n_{3}+n_{4})F(1,1,n_{3},n_{4}-1,n_{5})$\\$(1-\D+n_{3}+n_{5})F(1,1,n_{3}-1,n_{4},n_{5})$\\$(1-\D+n_{3}+n_{5})F(1,1,n_{3},n_{4},n_{5}-1)$}&$F(1,1,0,0,0)$\\
        \hline
    \end{tabular}
\end{center}
The irreducible integrals table describes what integrals we still need to find reduction rules for in the next step.

\paragraph{Row Reducing Identities}

The table below shows what identities we manage to obtain after row reduction on each sector that still had unresolved integrals
\begin{center}
    \begin{tabular}{|c|c|c|}
        \hline
        Sector&Used Identities (leading weight terms)&Irreducible Integrals\\
        \hline
        $(2,1,1,0,0)$&$12(\D-5)(\D-1)m^{2}s(9m^{4}-10m^{2}s+s^{2})F(2,1,1,n_{4},0)$&N/A\\
        \hline
        $(1,1,1,0,0)$&N/A&\makecell{$F(1,1,1,0,-1)$\\$F(1,1,1,n_{4},0)$}\\
        \hline
        $(1,1,0,0,0)$&N/A&$F(1,1,0,0,0)$\\
        \hline
    \end{tabular}
\end{center} 

\paragraph{Solving Directly for Missing Rules}

The table below shows the reduction rules we found for each sector that still had unreduced integrals and the seed integrals we needed to insert into the identities to get them
\begin{center}
    \begin{tabular}{|c|c|c|c|}
        \hline
        Sector&Integral Solved&Input Seeds&Irreducible Integrals\\
        \hline
        $(1,1,1,0,0)$&$F(1,1,1,n_{4}-2,0)$&\makecell{$(1,1,1,n_{4},0)$\\$(1,1,1,n_{4}-1,0)$}&\makecell{$F(1,1,1,-2,0)$\\$F(1,1,1,0,-1)$\\$F(1,1,1,-1,0)$\\$F(1,1,1,0,0)$}\\
        \hline
        $(1,1,0,0,0)$&N/A&\makecell{$(1,1,0,0,0)$\\$(1,1,0,0,-1)$\\$(1,1,0,-1,0)$\\$(1,1,-1,0,0)$}&$F(1,1,0,0,0)$\\
        \hline
    \end{tabular}
\end{center}
For sector $(1,1,1,0,0)$, we only needed to input two generic seeds to find a generic reduction rule for $F(1,1,1,n_{4}-2,0)$, however we found that this rule was invalid when $n_{4}=0$, as it is of the form
\begin{equation}
    F(1,1,1,n_{4}-2,0)\rightarrow\frac{(\cdots)}{4(3\D-4-2n_{4})n_{4}^{2}}\,.
\end{equation}
We are left with $4$ irreducible integrals in this sector, which are our master integrals. For sector $(1,1,0,0,0)$ we input a few seeds around this point however we cannot reduce the integral $F(1,1,0,0,0)$, therefore this is also a master integral.

\paragraph{Applying Reduction Rules}

In order to apply the reduction rules we follow the flowchart in \ref{fig:equations}. We start with a queue of the following integrals
\begin{equation}
    F(2,1,1,-4,-4), F(1,2,1,0,-7), F(1,1,1,-6,-4), F(1,1,1,-11,0)
\end{equation}
We start by taking the highest weight integral $F(2,1,1,-4,-4)$ and we generate an equation that acts as a reduction rule. In this case the equation is
\begin{equation*}
    \begin{aligned}
    0&=-6 F(2,1,1,-5,-2) \left(m^2+s\right)+F(2,1,1,-4,-3) \left(m^2+3 s\right)+3 s F(2,1,1,-4,-2) \left(m^2+s\right)\\
    &\quad+8 F(2,1,1,-3,-4) \left(m^2+s\right)-4 s F(2,1,1,-3,-3) \left(m^2+s\right)+3 s F(1,1,1,-4,-2)\\
    &\quad-4 s F(1,1,1,-3,-3)+3 s F(2,0,1,-4,-2)-4 s F(2,0,1,-3,-3)-3 s F(2,1,0,-4,-2)\\
    &\quad+4 s F(2,1,0,-3,-3)+F(1,1,1,-4,-3)+8 F(1,1,1,-3,-4)-6 F(2,0,1,-5,-2)\\
    &\quad+F(2,0,1,-4,-3)-F(2,1,0,-4,-3)+4 F(2,1,1,-5,-3)-10 F(2,1,1,-4,-4)
    \end{aligned}
\end{equation*}
This is an equation where $F(2,1,1,-4,-4)$ is the highest weight integral. We then add to the queue the new integrals appearing in this equation, so the queue now contains $20$ integrals. We also add this equation to our list of equations and add the integral $F(2,1,1,-4,-4)$ to our list of reduced integrals. We then move on to the next highest weight integral of queue, which will now be $F(2,1,1,-5,-3)$ and generate a similar equation for this integral. We do this iteratively until our queue is empty. For some integrals, there is no reduction rule for it, so we add it to the list of master integrals. Doing this iteratively we find a system of $690$ equations in 0.23s, which we can then perform backward substitution on to find the reduction.

\bibliographystyle{JHEP}
\bibliography{biblio}

\end{fmffile}

\end{document}